\documentstyle[11pt]{article}

    \advance\textheight by \topskip

\begin{document}
\hfill
UM-TH-97-02

\hfill
NIKHEF-97-011

\begin{center}
{\Huge \bf The $\alpha_s^3$ approximation of Quantum
Chromodynamics to the Ellis-Jaffe sum rule} \\[8mm]
  S.A. Larin$^{a}$, T. van Ritbergen$^b$, J.A.M. Vermaseren$^c$ \\ [3mm]
\begin{itemize}
\item[$^a$]
 Theory Division, CERN, CH-1211, Geneva 23, Switzerland\\
 and Institute for Nuclear Research of the
 Russian Academy of Sciences,   \\
 60th October Anniversary Prospect 7a,
 Moscow 117312, Russia
\item[$^b$]
 Randall Laboratory of Physics, University of Michigan,\\
 Ann Arbor, MI 48109, USA
\item[$^c$]
 NIKHEF, P.O. Box 41882, \\ 1009 DB, Amsterdam, The Netherlands \\
\end{itemize}
\end{center}

\begin{abstract}
We present the analytical calculation in perturbative
Quantum Chromodynamics of the $\alpha_s^3$
contribution to the Ellis-Jaffe sum rule for the structure function $g_1$
of polarized deep inelastic lepton-nucleon scattering.  \vspace{3cm} 
\end{abstract}
\newpage

Measurements of the polarized nucleon structure function
$g_1$ during the last 20 years have revealed an internal spin structure
of the nucleon that is surprisingly different from constituent quark model
expectations.
The discovery of the disagreement between the combined 
EMC-SLAC data \cite{ashman,slacdata} 
and the constituent quark model expectation \cite{ejfirst}
$\int_0^1 dx g_1^p(x,Q^2) \approx 0.15 |g_A|$  attracted a lot of attention
and triggered intensive research in the field of polarized deep inelastic
scattering.
More recently, deuterium scattering data 
of the SMC \cite{smc} and
of the E143 \cite{E143} collaborations,
and $^3$He scattering data of the E142 \cite{E142} 
and E154 \cite{E154prime} collaborations also allowed the 
determination of the neutron sum rule
$\int_0^1 dx g_1^n(x,Q^2)$ and the Bjorken sum rule \cite{bjorkenfirst}
$\int_0^1 dx [g_1^p(x,Q^2)- g_1^n(x,Q^2)]$.
At present the Bjorken sum rule is confirmed at the 8\% level 
\cite{testbjorken} which is an important experimental test of
Quantum Chromodynamics.

Higher order perturbative QCD corrections to the sum rules are crucial for 
an accurate and reliable confrontation of these sum rules with 
experimental data, see e.g. Ref. \cite{relevance}. 
For the Bjorken sum rule, the $\alpha_s$ correction \cite{bj1loop},
the $\alpha_s^2$ correction \cite{bj2loop},
and the $\alpha_s^3$ correction \cite{bj3loop}
have been calculated in the leading twist approximation. 
Higher twist corrections have also been calculated \cite{highertwists}. 
The Ellis-Jaffe sum rule $\int_0^1 dx g_1^{p/n}(x,Q^2)$ 
for the proton and neutron was calculated to order
$\alpha_s$ \cite{kod} and to order $\alpha_s^2$
\cite{ej2loop} in the leading twist approximation. 
Power corrections were calculated in \cite{powercorrections}.

In this article we obtain the order $\alpha_s^3$ contribution
to the Ellis-Jaffe sum rule in the leading twist approximation
for massless quarks.
We perform the calculations using dimensional regularization
\cite{dimreg} in
$D=4-2\varepsilon$ space-time dimensions and use the standard modification
of the minimal subtraction scheme \cite{ms}, the $\overline{\rm MS}$-scheme 
\cite{msbar}.

Polarized deep inelastic electron-nucleon scattering is described
by the hadronic tensor
\begin{eqnarray}
 W_{\mu\nu} & = & \frac{1}{4\pi} \int d^4z e^{iqz} \langle p,s|
 J_\mu(z) J_{\nu}(0) |p,s\rangle \nonumber \\
 & = & \left( -g_{\mu\nu} +\frac{q_\mu q_\nu}{q^2} \right) F_1(x,Q^2)
   + \left( p_\mu-\frac{p\cdot q}{q^2} q_\mu \right) 
       \left( p_\nu - \frac{p\cdot q}{q^2} q_\nu \right) 
      \frac{1}{p\cdot q} F_2(x,Q^2)  \nonumber \\
 & & + i \epsilon_{\mu\nu\rho\sigma} q_{\rho} \left(
        \frac{ s_\sigma}{p\cdot q} g_1(x,Q^2)
         +\frac{s_\sigma p\cdot q - p_\sigma q\cdot s}{ (p\cdot q)^2}
         g_2(x,Q^2) \right)
\end{eqnarray}
Here $J_\mu = \sum_{i=1}^{n_f} e_i \overline{\psi_i} \gamma_\mu \psi_i$ is the 
electromagnetic quark current where $ e_i =2/3,-1/3,-1/3,\cdots $ 
is the electromagnetic charge of a quark with the corresponding flavour.
$x=Q^2/(2p\cdot q)$ is the Bjorken scaling variable and $Q^2 = -q^2$ is the
square of the transferred momentum. $ |p,s\rangle $ is the
nucleon state that is normalized as $\langle p,s | p', s' \rangle$ 
$= 2 p^0 (2\pi)^3 \delta^{(3)}(p-p') \delta_{ss'}$. The polarization vector
of the nucleon is expressed as 
$s_\sigma = \overline{U}(p,s) \gamma_\sigma \gamma_5 U(p,s)$ where $U(p,s)$
is the nucleon spinor $\overline{U}(p,s) U(p,s) = 2 M$.

In the present article we will focus on the first Mellin moment of the
structure function $g_1$. Moments of deep inelastic structure functions 
can be expressed \cite{chm} (for reviews see Refs. \cite{buras,reya})
in terms of quantities that appear in the operator product
expansion (OPE) of the two currents $J_\mu$. 
For the first moment of the structure function $g_1$ we need to consider the
following expression for the OPE of two electromagnetic currents

\begin{eqnarray} 
 i\int dz e^{iqz} T \{ J_\mu(z) J_\nu (0)\}  
 & \stackrel{Q^2 \rightarrow \infty}{=} 
 \epsilon_{\mu\nu\rho\sigma} \frac{q_\rho}{q^2} \Biggl[ &
\sum_a C^a \left(\log (\frac{\mu^2}{Q^2}),a_s(\mu^2) \right)J_\sigma^{5,a} (0) 
  \nonumber \\ & &
  +C^s \left(\log (\frac{\mu^2}{Q^2} ),a_s(\mu^2) \right)J_\sigma^{5} (0)
 \Biggr] + \cdots \label{ope} 
\end{eqnarray}
where everything is assumed to be renormalized (with $\mu$ being the
renormalization scale). $J_\sigma^{5,a}(x) = \overline{\psi} \gamma_\sigma
 \gamma_5 t^{a} \psi(x)$ is the non-singlet axial current, where $t^a$ is a
generator of the flavour group, and $J_\sigma^5(x) =
  \sum_{i=1}^{n_f} \overline{\psi_i} \gamma_\sigma \gamma_5 \psi_i(x)$
is the singlet axial current.
We use the notation
 \begin{equation} \label{defineas}
 a_s(\mu^2) = \frac{g^2(\mu^2)}{16\pi^2} = \frac{ \alpha_s(\mu^2)}{4\pi}
\end{equation}
for the QCD strong coupling constant, where $g(\mu^2)$ is the
coupling constant of the QCD Lagrangian. 

 The dots on the r.h.s of Eq. (\ref{ope}) indicate 
higher spin and higher twist contributions that do not contribute to the
first moment of $g_1$ in the leading twist approximation.
The strict method of the OPE ensures \cite{collins} that the OPE of 
two gauge-invariant
currents can only contain gauge invariant operators with their 
renormalization basis. For example, the known twist-two and spin-one 
axial gluon current $K_\sigma = 4 \epsilon_{\sigma\mu\nu\rho}
  (A^a_{\mu} \partial_{\nu} A^a_{\rho} + g f^{abc} A^a_\mu A^b_\nu A^c_\rho/3)$
also has the necessary quantum numbers, but it cannot contribute to the
above OPE because it is not gauge invariant. Since the formalism
of the OPE gives the strict theoretical basis to the parton model
with the parton distributions being the matrix elements
of the corresponding operators,  
there is strictly no well founded way to get the polarized
gluon distribution into the considered sum rule.

Since the non-singlet
coefficient function $C^a$ depends trivially on the flavour number $a$
(see e.g. Ref. \cite{momnonsinglet}) it is possible to factorize a 
flavour number independent coefficient
function for which we use the standard notation $C^{\rm ns}$.

 The Ellis-Jaffe sum-rule is expressed as 
\begin{eqnarray} \label{ejbasic}
\int_0^1 dx g_1^{p(n)} (x,Q^2) & = & C^{\rm ns}(1,a_s(Q^2))
  ( \pm \frac{1}{12}|g_A|
  +\frac{1}{36} a_8 ) \nonumber \\
 & & + C^{\rm s}(1,a_s(Q^2)) \exp\left( \int_{a_s(\mu^2)}^{a_s(Q^2)}
        da_s' \frac{\gamma^s(a_s')}{\beta(a_s')}\right)
                  \frac{1}{9} a_0(\mu^2)
\end{eqnarray}
where the plus (minus) sign before $|g_A|$ corresponds to the proton (neutron)
target. The proton matrix elements of the axial currents are
defined as  

\begin{equation}
\label{gaa8}
 \begin{array}{llrll}
 |g_A| s_\sigma & = & 2 \langle p,s | J^{5,3}_\sigma | p,s \rangle & = &
         (\Delta u - \Delta d ) s_\sigma , \nonumber \\
 a_8 s_\sigma & = & 2 \sqrt{3} \langle p,s | J^{5,8}_\sigma | p,s \rangle
 & = & ( \Delta u + \Delta d - 2 \Delta s) s_\sigma , \nonumber \\
 a_0(\mu^2)s_\sigma & = & \langle p,s | J^{5}_\sigma | p,s \rangle
 & = & ( \Delta u + \Delta d +  \Delta s) s_\sigma 
      = \Delta\Sigma (\mu^2) s_\sigma.
\end{array} 
\end{equation}
Here $|g_A|$ is the absolute value of the constant of 
the neutron beta-decay, $g_A/g_V = -1.2601 \pm 0.0025$ \cite{properties}.
$a_8=0.579 \pm 0.025$ \cite{properties,clro} is the constant of hyperon
decays. The matrix element of the singlet 
axial current $a_0(\mu^2)$ will be redefined in a proper
invariant way as a constant ${\hat a}_0$ below in Eq.(\ref{ainv}).

We use the  notation
$ \Delta q(\mu^2) s_\sigma = \langle p,s | \overline{q} \gamma_{\sigma}
 \gamma_5 q |p,s \rangle$, $q = u,d,s$,
for the polarized quark distributions.
We omit the contributions
of the nucleon matrix elements for quarks heavier than the s-quark
but it is straightforward to include them. We also avoid
to introduce the contribution
of the polarized gluon distribution $\Delta g$ in the expression for
the matrix element of the singlet axial current $a_0(\mu^2)$ 
for the reason given below Eq. (\ref{defineas}).

$\beta(a_s)$ is the beta function that determines the
renormalization scale dependence of the renormalized coupling constant.
It is presently known at four loops \cite{beta4loops}
in the  $\overline{\rm MS}$ scheme
\begin{eqnarray}
  \beta (a_s)  & = & \frac{\partial a_s }{\partial \, \ln \mu^2}   
 =   -\beta_0 a_s^2 - \beta_1 a_s^3 -\beta_2 a_s^4
      -\beta_3 a_s^5 + O(a_s^6)  \nonumber
\end{eqnarray}
with the SU(3) values
\begin{eqnarray}
\renewcommand{\arraystretch}{ 1.3} 
 \beta_0 & = &  \textstyle 11 - \frac{2}{3} n_f   \nonumber\\
 \beta_1 & = &\textstyle  102 - \frac{38}{3} n_f  \nonumber \\
 \beta_2 & = &  \textstyle \frac{2857}{2} - \frac{5033}{18} n_f + \frac{325}{54}
  n_f^2 \nonumber \\
 \beta_3 & = & \textstyle  \left( \frac{149753}{6} + 3564 \zeta_3 \right)
        - \left( \frac{1078361}{162} + \frac{6508}{27} \zeta_3 \right) n_f
  \nonumber \\ & & \textstyle
       + \left( \frac{50065}{162} + \frac{6472}{81} \zeta_3 \right) n_f^2
       +  \frac{1093}{729}  n_f^3
\end{eqnarray} 
in which $\zeta$ is the Riemann zeta-function 
($\zeta_3 = 1.202056903\cdots$).

The singlet anomalous dimension $\gamma^{\rm s}$ that appears 
in Eq. (\ref{ejbasic})
determines the renormalization
scale dependence of the axial singlet current $J^5_\sigma$,
\begin{equation}
\frac{ d}{d\ln \mu^2} [J^5_{\sigma}]_R = \gamma^{\rm s} [J^5_{\sigma}]_R
\end{equation}
where subscript $R$ means that a current is renormalized.

The axial singlet current is not conserved -- the axial anomaly 
\cite{axialanomaly} -- and this causes
the singlet anomalous dimension to be non-zero starting from the order $a_s^2$ 
\begin{equation}
 \gamma^{\rm s}(a_s) = \gamma_1 a_s^2 + \gamma_2 a_s^3 + \gamma_3 a_s^4 + O(a_s^5).
\end{equation}
That is why $a_0(\mu^2)$ in Eq. (\ref{ejbasic}) depends on the 
renormalization point $\mu^2$ and it is therefore not a physical quantity.

The {\em non-singlet} axial current is conserved in the limit
of massless quarks and the anomalous dimension for the non-singlet 
axial current therefore vanishes. 
This is why the non-singlet contribution to the Ellis-Jaffe
sum rule Eq. (\ref{ejbasic}) does not involve an exponential factor
and that is why the non-singlet matrix elements $g_A$ and $a_8$ are 
renormalization group invariant ($\mu^2$ independent).

The non-singlet contribution to the Ellis-Jaffe sum rule is known in 
the order $a_s^3$ from \cite{bj3loop} where the polarized Bjorken 
sum rule $\int_0^1 dx (g_1^p - g_1^n)$ was calculated in this order. 
In this article we obtain the order $a_s^3$ contribution to the singlet part 
of the Ellis Jaffe sum rule:

\[ C^{\rm s}(1,a_s(Q^2)) \exp \left[ \int_{a_s(\mu^2)}^{a_s(Q^2)}
        da_s' \frac{\gamma^{\rm s}(a_s')}{\beta(a_s')}\right]
                  \frac{1}{9}a_0(\mu^2)  \hspace{7cm}\]
\[ =  C^{\rm s}(1,a_s(Q^2)) \left[ 
       1+ a_s(Q^2)\frac{-\gamma_1}{\beta_0}  
  + (a_s(Q^2))^2 \frac{ -\gamma_2 \beta_0
          +\gamma_1 \beta_1+\gamma_1^2}{2 \beta_0^2}  \right. \hspace{3.3cm} \]
\begin{equation} \left.
  + (a_s(Q^2))^3 
    \frac{ -2 \gamma_3 \beta_0^2 
        +2 \gamma_1 \beta_2 \beta_0+2 \beta_1 \gamma_2 \beta_0 
     -2 \gamma_1 \beta_1^2
     +3 \gamma_1 \gamma_2 \beta_0 
     -3 \gamma_1^2 \beta_1 - \gamma_1^3}{ 6 \beta_0^3 }  \right]
         \frac{1}{9} {\hat a}_0 \label{ejinvariant}
\end{equation}
where we introduce the notation
\begin{equation}
\label{ainv}
 {\hat a}_0 = \exp \left( - \int^{a_s(\mu^2)}
        da_s' \frac{\gamma^s(a_s')}{\beta(a_s')}\right) a_0(\mu^2) 
\end{equation}
for the renormalization group invariant (i.e. $\mu^2$ independent) nucleon
matrix element of the singlet axial current. 
Since ${\hat a}_0$
is the renormalization group invariant it should be considered as
a physical constant on the same ground as the constants $g_A$ and $a_8$.
That is why from now on we will only use the notation ${\hat a}_0$
for the matrix element of the singlet axial current in the
expression for the sum rule.

To obtain this singlet contribution, we need to calculate the 3-loop 
contribution to the singlet coefficient function $C^{\rm s}$ and 
the 4-loop contribution to the anomalous dimension $\gamma^{\rm s}$.
For the treatment of the $\gamma_5$ matrix in dimensional regularization
we use a technique described in \cite{gamma5technique} which is based on 
the original definition of $\gamma_5$ in \cite{dimreg}.

In the $\overline{\rm MS}$ scheme the proper normalization of the 
axial singlet current requires the
introduction of a finite renormalization constant $Z^{\rm s}_5$ in addition
to the standard ultraviolet renormalization constant\footnote{
Please note that  ultraviolet renormalization constants coincide in the MS 
and $\overline{\rm MS}$ schemes and we can therefore use the notation 
$Z_{\rm MS}$ instead of $Z_{\overline{\rm MS}}$.}
 $Z^{\rm s}_{\rm MS}$ that
contains only poles in the regularization parameter $\varepsilon$
\begin{equation} \label{defZ5}
 [j_{\mu}^5]_R = Z^{\rm s}_5 Z^{\rm s}_{\rm MS} [J_\mu^5]_B 
 = Z^{\rm s}_5 Z^{\rm s}_{\rm MS}
 \frac{i}{6} \epsilon_{\mu\nu\rho\sigma} \overline{\psi}_B
   \gamma_{\nu} \gamma_{\rho} \gamma_{\sigma} \psi_B 
\end{equation}
where $B$ denotes the bare, unrenormalized quantity. 

This extra renormalization constant is introduced to keep 
the exact 1-loop Adler-Bardeen form \cite{adlerbardeenform}
for the operator anomaly 
equation $\partial_\mu J_\mu^5 =
 a_s \frac{n_f}{2} \epsilon_{\mu\nu\lambda\rho}G^a_{\mu\nu}G^a_{\lambda\rho}$
within dimensional regularization.
Here $G^a_{\mu\nu} = \partial_\mu A_\nu^a -\partial_\nu A_\mu^a
     +ig f^{abc} A^b_\mu A^c_\nu$ is the QCD field strength tensor.

This finite renormalization constant $Z_5^{\rm s}$ 
affects the results for the coefficient function 
and anomalous dimension but the sum rule, i.e. the combination in the r.h.s. of 
Eq. (\ref{ejinvariant}) is independent of the choice of $Z_5^{\rm s}$. 
This fact is evident since the sum rule as a physical object
can not depend of the choice of the normalization of a
non-physical object such as a flavour singlet axial current.
More precisely, $Z_5^{\rm s}$ enters in the coefficient function 
and anomalous dimension as
\begin{eqnarray}
 C^{\rm s}  & = & \overline{C^{\rm s}} / Z_5^{\rm s},\nonumber \\ 
 \label{transform}
  \gamma^{\rm s} & = & \overline{\gamma^{\rm s}}
   + \beta(a_s) \frac{d}{d a_s} log(Z_5^{\rm s}) 
\end{eqnarray}
where $\overline{C^{\rm s}}$ and $\overline{\gamma^{\rm s}}$ are calculated
in the $\overline{\rm MS}$ scheme without a factor $Z_5^{\rm s}$ 
(i.e. with $Z_5^{\rm s} = 1$). Please notice that $Z_5^{\rm s}$ 
enters the anomalous
dimension only at the $a_s^2$ order, since $\beta (a_s)$ starts from $a_s^2$.
Presently the singlet constant $Z_5^{\rm s}$ is unknown in the order $a_s^3$.
We will therefore obtain $\overline{C^{\rm s}}$ and 
 $\overline{\gamma^{\rm s}}$ which is sufficient to obtain the sum rule.

To obtain the 3-loop coefficient function $\overline{C^{\rm s}}$ 
we  apply the method of projectors \cite{projectors}
which gives us the following formula 
\begin{equation} \label{basicprojector}
  \overline{C^{\rm s}} = \frac{1}{24} \frac{1}{Z^{\rm s}_{\rm MS}} 
 R_{\overline{\rm MS}}
 \,  i \int dz e^{i qz} \langle 0 | T \{ \overline{\psi}(p) \gamma_{[\mu}
 \gamma_{\nu} \gamma_{\rho ]} q_\rho \psi(-p) J_\mu (z) J_\nu (0) \}
   |0\rangle |^{\rm amputated}_{p=0}
\end{equation}
where $ \gamma_{[\mu} \gamma_{\nu} \gamma_{\rho ]} q_\rho
      = \gamma_\mu \gamma_\nu \!\! \not \! q -\delta_{\mu\nu} \!\! \not \! q
        +q_\mu \gamma_\nu - q_\nu \gamma_\mu $.
The $R_{\overline{\rm MS}}$ operation performs the renormalization of
the ultraviolet divergences. The infrared divergences that are 
produced by putting $p=0$ are removed by the ultraviolet 
renormalization factor of the axial singlet current $Z^{\rm s}_{\rm MS}$.
In this way we need to evaluate 3-loop massless propagator diagrams 
which is done with the package MINCER written for the symbolic manipulation
program FORM \cite{form}.
The set of 3-loop flavour singlet diagrams that contribute 
to this coefficient function is identical to the set ``q$\gamma$q$\gamma$" 
that previously appeared in Ref. \cite{momsin}.
In this way we obtained the following result for $\overline{C^{\rm s}}$
\begin{eqnarray}
\overline{C^{\rm s}} & = &  1 
      + a_s C_F  ( - 7) \nonumber \\ & &
       + a_s^2 \Bigl[ C_A C_F  ( - \frac{314}{9} )
       + C_F n_f T_F (  \frac{109}{9} +  16 \zeta_3 )
       + C_F^2 ( \frac{89}{2} ) \Bigr] \nonumber \\ & &
     + a_s^3 \Bigl[ C_A C_F n_f T_F (  \frac{41215}{81} + \frac{1124}{9}\zeta_3)
       + C_A C_F^2  (  \frac{13004}{27} - \frac{656}{3} \zeta_3)
         \nonumber \\ & & \hspace{5mm} 
       + C_A^2 C_F  ( - \frac{13021}{27} + 56 \zeta_3 + \frac{440}{3} \zeta_5)
       + C_F n_f^2 T_F^2  ( - \frac{9404}{81} - \frac{256}{9} \zeta_3)
         \nonumber \\ & & \hspace{5mm} 
       + C_F^2 n_f T_F ( - \frac{460}{27} - 320 \zeta_5)
       + C_F^3 ( - \frac{1397}{6}  + 96 \zeta_3 ) \Bigr] \label{rescoef}
\end{eqnarray}
where $C_F=4/3$ and $C_A=3$ are the quadratic
 Casimir operators of the fundamental and
adjoint representation of the colour group SU(3), $T_F=1/2$ is the 
trace normalization of the fundamental representation and $n_f$ is the
number of (active) quark flavours. The Riemann zeta function is written as 
$\zeta_n$. 
It is interesting to mention
that diagrams with one external photon in a closed
quark loop have colour factors proportional to the cubic 
Casimir operator $d^{abc}d^{abc}$. Individually these diagrams are non-zero
but this higher qroup invariant cancels in the sum.


To obtain the 4-loop contribution to the anomalous dimension 
$\overline{\gamma^{\rm s}}$ we need to calculate the  
$\overline{\rm MS}$ renormalization factor $Z_{\rm MS}^{\rm s}$
 of Eq. (\ref{defZ5}) 
(and of Eq. (\ref{basicprojector}) where the lower order $Z_{\rm MS}^{\rm s}$
 appeared)
which contains apart from a leading constant 1 only poles in the
regularization parameter $\varepsilon$,
\begin{equation}
 Z_{\rm MS}^{\rm s}(a_s,\varepsilon) 
= 1+Z^{{\rm s},(1)}(a_s)/\varepsilon + Z^{{\rm s},(2)}(a_s)/\varepsilon^2 
  + \cdots
\end{equation}
The anomalous dimension $\overline{\gamma^{\rm s}}$ is then expressed through 
the coefficient in front of the first pole in $Z_{\rm MS}^{\rm s}$
\begin{equation}
 \overline{\gamma^{\rm s}} = - a_s \left( \frac{\partial}{\partial a_s}
  Z^{{\rm s},(1)}(a_s)
 \right)
\end{equation}
To obtain the renormalization factor $Z_{\rm MS}^{\rm s}$ we 
 calculated the overall ultraviolet
divergence of the Green function
\begin{equation} G_{\overline{\psi}[J_\mu^5]\psi} =
 \int dx dy e^{i qx + ipy} 
  \langle 0 | T\{ \overline{\psi}(x) J_\mu^5(y)  \psi(0) \} |0 \rangle
\end{equation}
This Green function is renormalized multiplicatively and the standard 
$\overline{\rm MS}$ renormalization factor (containing only poles
in $\varepsilon$) is equal to $Z_{\rm MS}^{\rm s}/Z_2$, where $Z_2$ is 
the renormalization factor of the inverted quark propagator.

To obtain the required anomalous dimension $\overline{\gamma^{\rm s}}$ 
in the $a_s^4$ order we calculated the overall divergences of both
the quark propagator and the Green function 
$G_{\overline{\psi}[J_\mu^5]\psi}$ in the 4-loop order. 
This can be conveniently done using the technique 
that is described in Ref. \cite{beta4loops}.
This general technique is based on the direct calculation of 4-loop massive 
vacuum (bubble) integrals and provides a procedure that is well suited for the
automatic evaluation of huge numbers of Feynman diagrams.
For the present calculation we needed to evaluate of the order of
 10000 4-loop diagrams. 
These diagrams were generated with the program QGRAF \cite{qgraf}.

\begin{eqnarray}
\overline{\gamma^{\rm s}} & = &  
       \hspace{3mm}  a_s^2 \Bigl[ 
          C_A C_F  ( - \frac{44}{3} )
         + C_F n_f T_F  ( - \frac{20}{3} )
        \Bigr] \nonumber \\ & &
     + a_s^3 \Bigl[ 
        C_A C_F^2 (   \frac{308}{3} )
       + C_A^2 C_F  ( -\frac{3578}{27} )
         \nonumber \\ & & \hspace{5mm}
       + C_A C_F n_f T_F ( -\frac{298}{27} )
       + C_F^2 n_f T_F (   \frac{44}{3} ) 
       + C_F n_f^2 T_F^2 ( -\frac{104}{27} )
        \Bigr] \nonumber \\ & &
     + a_s^4 \Bigl[ 
        C_A C_F^3  ( - \frac{1870}{3} + 1056 \zeta_3 )
       + C_A^2 C_F^2  (   \frac{58618}{27} - 1760 \zeta_3 )
         \nonumber \\ & & \hspace{5mm}
       + C_A^3 C_F ( - \frac{36607}{27} + 616 \zeta_3 )
       + C_A C_F^2 n_f T_F (   \frac{3794}{27} - \frac{368}{3} \zeta_3 )
         \nonumber \\ & & \hspace{5mm}
       + C_A^2 C_F n_f T_F  (   \frac{15593}{81} + \frac{1748}{3} \zeta_3 )
       + C_F^3 n_f T_F  ( - \frac{58}{3} - 384 \zeta_3 )
         \nonumber \\ & & \hspace{5mm}
       + C_F^2 n_f^2 T_F^2  (   \frac{6808}{27} - \frac{896}{3} \zeta_3 )
       + C_A C_F n_f^2 T_F^2  (   \frac{496}{81} + \frac{848}{3} \zeta_3 )
         \nonumber \\ & & \hspace{5mm}
       + C_F n_f^3 T_F^3  (   \frac{560}{81} )
        \Bigr] \label{resgamma}
\end{eqnarray}

The results of Eqs.(\ref{rescoef},\ref{resgamma}) are obtained in an arbitrary
covariant gauge for the gluon field.
This means that we keep
the gauge parameter $\xi$ that appears in the gluon propagator
$ i$  $ [-g^{\mu\nu}+(1-\xi)
 q^{\mu}q^{\nu}/(q^2+i\epsilon)]/(q^2+i\epsilon)$  as a
free parameter in the calculations. The explicit cancellation of the gauge
dependence in the coefficient function  and the anomalous dimension gives 
an important check of the results.
The results for individual
diagrams that contribute to $\overline{C^{\rm s}}$ and 
 $\overline{\gamma^{\rm s}}$ also contain (apart from the constant $\zeta_3$) 
the constants $\zeta_4$, $\zeta_5$.
The cancellation of these constants at various stages in the calculation
provides additional checks of the results.
We should also note that the various higher order colour factors 
[see Ref. \cite{beta4loops}] 
that appear in the separate results for $Z_2$ and $(Z^{\rm s}_{\rm MS}/Z_2)$
all canceled in Eq. (\ref{resgamma}).

At this point it is interesting to compare the obtained result for 
$\overline{\gamma^{\rm s}}$ with the analogously 
defined {\em non-singlet} anomalous
dimension $\overline{\gamma^{\rm ns}}$ since this can be constructed at 4-loops
from known 3-loop results. Since the non-singlet anomalous dimension
$\gamma^{\rm ns}$ vanishes, we have [see, e.g. Eq. (\ref{transform})]
$\overline{\gamma^{\rm ns}} = - \beta(a_s) \frac{d}{d a_s}
  log(Z_5^{\rm ns})$ where
$Z_5^{\rm ns}$ is the finite renormalization constant for the
non-singlet axial current which is known in the order $a_s^3$.
Since the difference between the singlet and non-singlet sector is in
terms of at least one power of $n_f$ 
we must have that all the $n_f$-independent terms in 
 $\overline{\gamma^{\rm ns}}$
and $\overline{\gamma^{\rm s}}$ coincide, and indeed they do.
This gives another strong check to the calculations.

Substitution of the obtained anomalous dimension and coefficient function
in Eq. (\ref{ejinvariant}) gives the following result for the Ellis-Jaffe
sum rule.

\[ \int_0^1 dx g_1^{p(n)} (x,Q^2) =  \left[ 1 
 + \left( \frac{\alpha_s}{\pi} \right) d^{\rm ns}_1
 + \left( \frac{\alpha_s}{\pi} \right)^2 d^{\rm ns}_2
 + \left( \frac{\alpha_s}{\pi} \right)^3  d^{\rm ns}_3 \right] 
     (\pm \frac{1}{12}|g_A| + \frac{1}{36} a_8 ) \]
\[    \hspace{2cm} + \left[ 1 
 + \left( \frac{\alpha_s}{\pi} \right) d^{\rm s}_1
 + \left( \frac{\alpha_s}{\pi} \right)^2 d^{\rm s}_2
 + \left( \frac{\alpha_s}{\pi} \right)^3  d^{\rm s}_3 \right] 
      \frac{1}{9} {\hat a}_0 \]
\begin{eqnarray} 
d_1^{\rm ns} & = & -1 
   \nonumber \\
d_2^{\rm ns} & = & \textstyle  -\frac{55}{12} + \frac{1}{3} n_f
   \nonumber \\
d_3^{\rm ns} & = &
   \textstyle ( -\frac{13841}{216} -\frac{44}{9}\zeta_3 +\frac{55}{2}\zeta_5 )
     +n_f ( \frac{10339}{1296} + \frac{61}{54}\zeta_3 -\frac{5}{3}\zeta_5 )
     +n_f^2 (-\frac{115}{648} )
   \nonumber \\
d_1^{\rm s\;\;} & = & (1/\beta_0) \textstyle \Bigl[
           - 11 
      + n_f  ( \frac{8}{3} ) \Bigr]
   \nonumber \\
d_2^{\rm s\;\;} & = &  (1/\beta_0)^2  \textstyle \Bigl[
          - \frac{6655}{12} 
       + n_f  ( \frac{235}{2} + \frac{242}{3} \zeta_3 )
       + n_f^2  (  - \frac{85}{18} - \frac{88}{9} \zeta_3 )
       + n_f^3  ( \frac{16}{81} + \frac{8}{27} \zeta_3 ) \Bigr]
   \nonumber \\
d_3^{\rm s\;\;} & = & (1/\beta_0)^3  \textstyle \Bigl[
       (  - \frac{18422371}{216} - \frac{58564}{9} \zeta_3 
           + \frac{73205}{2} \zeta_5 )
       + n_f  ( \frac{46351373}{1296} + \frac{312785}{54} \zeta_3
                  - \frac{113135}{9} \zeta_5 ) 
   \nonumber  \\ & & \textstyle
   \hspace{.5cm}    + n_f^2  (  - \frac{2353243}{432} - \frac{30976}{27}\zeta_3
                 + \frac{13310}{9} \zeta_5)
       + n_f^3  ( \frac{4647815}{11664} + \frac{22594}{243} \zeta_3
                   - \frac{220}{3} \zeta_5 )
   \nonumber  \\ & & \textstyle
 \hspace{.5cm}   + n_f^4  (  - \frac{235867}{17496} - \frac{2440}{729} \zeta_3
               + \frac{320}{243}\zeta_5 )
       + n_f^5  ( \frac{386}{2187} + \frac{32}{729} \zeta_3 ) \Bigr]
 \label{result} \end{eqnarray}
where $\alpha_s = \alpha_s(Q^2)=4 \pi a_s(Q^2)$, 
$\beta_0 = 11 -2/3 n_f$ is the 
1-loop coefficient of the beta function and ${\hat a}_0$ is
the invariant matrix element of the singlet axial current defined
in Eq. (\ref{ainv}).

In particular, for $n_f=3$ we find
\[ \int_0^1 dx g_1^{p(n)} (x,Q^2) = \left[ 1
 - \left( \frac{\alpha_s}{\pi} \right) 
 -3.5833 \left( \frac{\alpha_s}{\pi} \right)^2 
 -20.2153 \left( \frac{\alpha_s}{\pi} \right)^3  \right]
     (\pm \frac{1}{12}|g_A| + \frac{1}{36} a_8 ) \]
\begin{equation}   \hspace{3.5cm} + \left[ 1
 - 0.33333 \left( \frac{\alpha_s}{\pi} \right) 
 - 0.54959 \left( \frac{\alpha_s}{\pi} \right)^2 
 - 4.44725 \left( \frac{\alpha_s}{\pi} \right)^3 \right]
      \frac{1}{9} {\hat a}_0 
\end{equation}

It is interesting to compare our result Eq. (\ref{result})  
with a recent estimate \cite{ej3loopprediction}
of the singlet coefficient $d_3^{\rm s}$ for $n_f=3$.
The estimate gives the value -2  which is 
slightly less than half the value obtained in the present article. 

In table 1 we have listed the numerical values of the second and third-order 
coefficients for the Ellis-Jaffe sum rule for $n_f = 3,4,5,6$.
One can observe the sign-constant character of perturbative QCD series
both for non-singlet and singlet contributions.
The series tends to preserve its sign-constant character
even when perturbative coefficients of the singlet contribution
change their signs around the value $n_f=4$.

\renewcommand{\arraystretch}{1.5}

\begin{center}
\begin{tabular}{crcrcrcrr}
\hline \hline 
 \,\,\,\,\,\,\,\, $n_f$ \,\,\,\,\,\,\,\, & \multicolumn{3}{c}{non-singlet} &  &
        \multicolumn{3}{c}{singlet} & \,\,\,\,\,\,\,\, \\ 
\cline{2-4} \cline{6-8}
 & \, $(\alpha_s/\pi)^2$ & \,\,\,\,\,&  \,$(\alpha_s/\pi)^3$  & 
      \,\,\,\,\,\,\,\,\,
 & \, $(\alpha_s/\pi)^2$ & \,\,\,\,\,&  \,$(\alpha_s/\pi)^3$  & \\
\hline \renewcommand{\arraystretch}{1.0}
3 &  -3.58333 & & -20.21527 & \,\,\, & -0.54959 & & -4.44725  & \\
4 &  -3.25000 & & -13.85026 & \,\,\, &  1.08153 & &  4.87423  & \\
5 &  -2.91667 & &  -7.84019 & \,\,\, &  2.97845 & & 13.07103  & \\
6 &  -2.58333 & &  -2.18506 & \,\,\, &  5.27932 & & 20.73034  & \\
\hline \hline
\end{tabular} \\
\vspace{5mm}
{\bf Table 1.} Second and third-order coefficients for the Ellis-Jaffe sum rule.
\end{center}

Another deep inelastic sum rule, the Bjorken sum rule for
neutrino-nucleon scattering, which is also known in the
$\alpha_s^3$-order \cite{bjorkenunpolarized}, also exibits the sign-constant
behaviour of the perturbative QCD series.
One can see that the obtained perturbative coefficients
of the Ellis-Jaffe sum rule grow rather moderately.
If we assume that the error of the truncated
asymptotic series is determined by the last calculated term, then the
obtained $\alpha_s^3$ approximation for this sum rule  provides a good
theoretical framework for extraction of the fundamental
constant ${\hat a}_0$ from experiment.

\section*{Acknowledgements}
We are grateful to J. Ellis, W.L. van Neerven, 
P.J. Nogueira and A.N. Schellekens 
for helpful and stimulating discussions. S.L. is grateful to the
Theory Division of CERN for its kind hospitality; his work is  
supported in part by the Russian Foundation
for Basic Research grant 96-01-01860.
The work of T.R. is supported by the US Department of Energy.

\end{document}